\documentclass[showpacs,nofootinbib,reprint,prl,aps,preprintnumbers]{revtex4-1}

\usepackage{amsmath,amssymb}
\usepackage{epsfig}  
\usepackage{graphicx}               
\usepackage{url}
\usepackage{hyperref}
\usepackage{bm}
\usepackage{tikz}
\usepackage[utf8]{inputenc}

\usepackage{pstricks}
\usepackage{color}
\usepackage{slashed}

\newcommand{\nc}{\newcommand}  

\newcommand{\mep}{m_{e'}}
\newcommand{\Minit}{M_\text{BH}^\text{init}}
\newcommand{\Mpl}{M_\text{pl}}

\nc{\beq}{\begin{equation}}  
\nc{\eeq}{\end{equation}}  
\nc{\beqa}{\begin{eqnarray}}  
\nc{\eeqa}{\end{eqnarray}}  
\nc{\bit}{\begin{itemize}}  
\nc{\eit}{\end{itemize}} 

\usepackage[normalem]{ulem} 

\begin{document}

\preprint{}

\title{Primordial Extremal Black Holes as Dark Matter
}
\author{Yang Bai and Nicholas Orlofsky} 
\affiliation{Department of Physics, University of Wisconsin-Madison,
  Madison, Wisconsin 53706, USA  
  }
\begin{abstract}
We show that primordial (nearly) extremal black holes with a wide range of  masses from the Planck scale to around $10^9$ g could be cosmologically stable and explain dark matter, given a dark electromagnetism and a heavy dark electron. For individual black holes, Hawking radiation and Schwinger discharge processes are suppressed by near-extremality and the heaviness of the dark electron, respectively. In contrast, the merger events of binary systems provide an opportunity to directly observe Hawking radiation.  Because the merger products are not extremal, they rapidly evaporate and produce transient high-energy neutrino and gamma ray signals that can be observed at telescopes like IceCube and HAWC. The relationship between the near-extremal black hole and dark electron masses could also shed light on the weak gravity conjecture.
\end{abstract}
\maketitle

\section{Introduction}
Primordial black holes (PBHs) \cite{Hawking:1971ei} are a compelling and simple explanation for dark matter (DM) \cite{Carr:1974nx,Carr:2016drx,Sasaki:2018dmp}.  They can take on a wide variety of masses, although several experiments constrain the allowable mass range~\cite{Carr:2009jm,Carr:2016drx,Niikura:2017zjd}, and there are proposals for future experiments to search much of the remaining parameter space~\cite{Katz:2018zrn,Bai:2018bej}.  Aside from these, perhaps the most important constraint is that they survive today.  Because black holes can evaporate via Hawking radiation \cite{Hawking:1974sw}, any Schwarzschild PBH with mass less than around $10^{15}$\,g has a lifetime less than the age of the universe and cannot account for DM.  A possible exception is that evaporation may cease when a black hole reaches the Planck mass, leaving behind a Planck-scale relic~\cite{Aharonov:1987tp} that could explain DM~\cite{MacGibbon:1987my}.  This is related to the information paradox~\cite{Hawking:1976ra}, and its status is still in debate~\cite{Chen:2014jwq}.  Even so, it would seem that PBHs cannot explain DM if they have mass anywhere between the Planck scale and $10^{15}~\text{g}$, a mass range we will call ``light.''

To stabilize such a light PBH against Hawking radiation, one could consider a Reissner-Nordstr\"om (RN) BH charged under an unbroken $U(1)$ gauge symmetry.  When the magnitude of the $U(1)$ charge $|Q|$ times the gauge coupling is equal to the BH mass in units of the Planck mass, the black hole temperature is zero and the Hawking evaporation does not occur.  Such a zero-temperature BH is called extremal~\footnote{More generally, a Kerr-Newman BH can become extremal by a combination of its charge and angular momentum, while a Kerr BH is uncharged and can become extremal by its angular momentum alone.  We do not consider these cases here because a non-extremal BH with angular momentum can lose angular momentum during its evaporation and will not approach to extremal when it evaporates \cite{Page:1976ki}.} (or eBH) and can be stable and serve as a DM candidate. 

Thus, it would seem as if giving a PBH some large Standard Model (SM) electric charge could allow it to survive today with mass higher than the Planck scale.  However, this is spoiled by the Schwinger effect~\cite{Schwinger:1951nm}, which allows pair-production of electron-positron pairs in the strong electric field outside of the eBH, leading to the eBH's discharge and subsequent evaporation~\cite{Gibbons:1975kk,Hiscock:1990ex,Sorkin:2001hf}. In this letter, to evade Schwinger discharge effects, we explore the possibility of having a stable primordial (nearly) extremal black hole (PeBH) charged under an unbroken dark $U(1)_{\rm dark}$ gauge symmetry, with its lightest charged state, or the ``dark electron," much heavier than the SM electron.

\section{Dark QED and extremal RN BH}
To achieve relatively light and extremal RN black holes, we make a simple assumption of physics beyond the SM in the dark sector: a dark quantum electromagnetic dynamics (dQED) with a dark electron with a mass $\mep$ and a gauge coupling $e'$
\beqa
\mathcal{L} \supset \mathcal{L}_{\rm SM} + \frac{1}{4} F'^{\mu\nu} F'_{\mu\nu} + \overline{e'} (i \gamma^\mu D_\mu - \mep) e'  ~,
\label{eq:dQED-lagrangian}
\eeqa
with $\mathcal{L}_{\rm SM}$ the SM Lagrangian, $D_\mu \equiv \partial_\mu + i e'A'_\mu$ the covariant derivative, and $F'_{\mu\nu} = \partial_\mu A'_\nu - \partial_\nu A'_\mu$. For simplicity, we assume that the dark electron is the lightest dQED charged particle and only include the dark electron in this study. At the renormalizable Lagrangian level, there is also an operator, $\epsilon\,F_{\mu\nu}F'^{\mu\nu}$, that kinetically mixes the visible photon ($F_{\mu \nu}$) and dark photon states, which would make the dark electron and PeBH millicharged under the visible photon. For simplicity, we neglect this term in the remainder of our analysis, reserving scrutiny of its implications for the discussion section.

The lower bound on a cosmologically long-lived eBH's mass is related to the particle physics model parameters in \eqref{eq:dQED-lagrangian}, $e'$ and $\mep$, via the non-perturbative Schwinger effect~\cite{Schwinger:1951nm}. More specifically, the pair production rate per unit volume from the Schwinger effect is~\cite{Schwinger:1951nm}
\beqa
\frac{d\Gamma_{\rm Schwinger} }{dV}= \frac{(e' E')^2}{4 \pi^3} \sum_{n=1}^{\infty} \frac{1}{n^2} \exp\left(-\frac{\pi\,n\, \mep^2}{e'\,E'}\right) \,,
\eeqa
with $E'$ as the dark electric field. For a RN eBH with $e' Q = \sqrt{4\pi} \, M_{\rm eBH} / \Mpl$ and $\Mpl = 1/\sqrt{G_N} = 1.22 \times 10^{19}$~GeV, the outer horizon radius is $r_+ = \left(M_{\rm eBH} + \sqrt{ M_{\rm eBH}^2 - e'^2 Q^2 \Mpl^2/4\pi }\right)/\Mpl^2 = M_{\rm eBH}/\Mpl^2$. The corresponding dark electric field just outside the horizon is $E' = \Mpl^3 / (\sqrt{4\pi}\,M_{\rm eBH})$. Note that $E'$ increases as $M_\text{eBH}$ decreases, so that the discharge speeds up as the PeBH discharges and evaporates. Requiring the evaporation time longer than the age of Universe, $t_{\rm univ}\approx 4.35\times 10^{17}$~s, one has
\beqa
M_{\rm eBH} > M^{\rm min}_{\rm eBH} \approx \frac{e' \Mpl^3}{2\pi^{3/2}\,\mep^2} \,\ln{ \left( \frac{e'^3\, \Mpl\,t_{\rm univ}}{16\, \pi^{7/2}} \right) }  \,.
\label{eq:Mmin}
\eeqa
Numerically, for $e' = e = \sqrt{4\pi \alpha}$, we have $M^{\rm min}_{\rm eBH} \approx 3.7 \times 10^{25}\,\mbox{GeV}\,\times \, (10^{16}\,\mbox{GeV}/\mep)^2$. Therefore, for a heavy dark electron mass or a small gauge coupling, the PeBH can be very light and even close to the Planck scale. Note that if we apply the formula in~\eqref{eq:Mmin} to SM QED with the electron mass $m_e =0.511$~MeV, then $M^{\rm min}_{\rm eBH} \approx 10^8\,M_\odot$---a mass that is already cosmologically stable without having to resort to an extremal charge. 

Na\"ivly, one can drive down the Schwinger discharge by arbitrarily decreasing $e'$. On the other hand, this is constrained by the so-called ``weak gravity conjecture" (WGC)~\cite{ArkaniHamed:2006dz,Brown:2015iha}, which suggests that $e' > \sqrt{4\pi}\,\mep/\Mpl$ for a particle with unit charge. Together with the WGC, we have a constraint on the eBH mass to be 
\beqa
M_{\rm eBH} 
 > M^{\rm min, WGC}_{\rm eBH} \approx \frac{\Mpl^2}{\pi\,\mep} \,\ln{ \left( \frac{\mep^3 \,t_{\rm univ}}{2\,\pi^{2}\, \Mpl^2} \right) }  \,.
\label{eq:Mmin-WGC}
\eeqa
Numerically, one has  $M^{\rm min, WGC}_{\rm eBH} \approx 2.9 \times 10^{23}\,\mbox{GeV} \times (10^{16}\,\mbox{GeV}/\mep)$. So, in principle, we could experimentally measure both $M_{\rm eBH}$ and $\mep$ to test the WGC. 

For an exactly extremal black hole, the Hawking temperature is zero. However, if the PeBH originally formed in a non-extremal state, it would take an infinite time to reach an extremal state. As a result, there is always a non-zero Hawking temperature for the PeBH. A general RN BH with a dark charge $Q$ has a Hawking temperature of
\beqa
T(M_{\rm BH}, M_{\rm eBH}) = \frac{\Mpl^2}{2\pi} \, \frac{ \sqrt{ M^2_{\rm BH}-  M_{\rm eBH}^2} }{ \left(M_{\rm BH} + \sqrt{ M^2_{\rm BH}-  M_{\rm eBH}^2 }\right)^2 }\,,
\eeqa
Using the black body radiation formula, the black hole mass loss rate is (see Ref.~\cite{Page:1976df} for spin-dependent corrections)
\beqa
\label{eq:dMdt}
\frac{dM_{\rm BH}}{dt} \approx - \frac{\pi^2}{120}\,g_*\, 4\pi r_+^2\, \left[T(M_{\rm BH}, M_{\rm eBH})\right]^4 \,,
\eeqa
with $g_*$ as the radiation degrees of freedom. When the initial black hole mass $\Minit \gg M_{\rm eBH}$ and ignoring the change of $g_*$, the black hole will evaporate to very nearly extremal with a lifetime $\tau_{\rm BH} \approx 5120\pi\,(\Minit)^3/(g_*\,\Mpl^4) \approx (1\,\text{s}) (\Minit/10^9\,\text{g})^3$.  After that, the black hole mass evolves with time as 
\beqa
M_{\rm BH}(t) 
=  M_{\rm eBH} + \frac{120\pi\,M^4_{\rm eBH}}{g_*\,\Mpl^4\,t} ~. 
\eeqa
During this later stage, the Hawking temperature of a nearly eBH has 
\beqa
T_{\rm eBH} = \sqrt{\frac{60\,M_{\rm eBH}}{\pi\,g_*\,t}}   ~.
\eeqa
Using $g_*\sim 10$ and the $t=t_{\rm univ}$, the temperature is $T_{\rm eBH} \approx 1.3\,\mbox{eV}\times \sqrt{M_{\rm eBH}/\mbox{g}}$. Generically, the evaporation from $\Minit$ to nearly extremal emits more Hawking radiation over a short period of time, while the emission of the near-extremal BH is suppressed but longer in duration.  As we will see, both these evaporation stages may provide observing opportunities.

\section{Searching for isolated $\mbox{PeBHs}$ }
Assuming PeBH's contribute to all or majority of the DM energy density, one could directly search for a PeBH in a laboratory. Using the whole Earth as a detector, the encounter rate of PeBH's is $\rho_{\rm DM} v_{\rm DM} \pi R_{\oplus}^2/M_{\rm eBH} \sim 1/\mbox{year}\times\left( 10^9\mbox{g}/M_{\rm eBH} \right)$ with $\rho_{\rm DM} \approx 0.4\,\mbox{GeV}/\mbox{cm}^3$ and $v_{\rm DM} \approx 10^{-3}\,c$. For a light PeBH, the geometric size of the PeBH is very small such that its interaction length for scattering off nucleons on the Earth is very long: $2\times 10^{11}\,\mbox{m}\times \left(10^9\,\mbox{g}/M_{\rm eBH} \right)^2$. So, traditional DM direct detection methods like measuring nuclear recoil energy are not sensitive to PeBHs. Instead, it may eventually be possible to measure the gravitational effects for a PeBH passing by mechanical resonators or free-falling masses, which may probe BH masses from the Planck mass scale to the gram level~\cite{Carney:2019pza}.

Rather than direct detection, given the fact that nearly-extremal BHs Hawking radiate photons, neutrinos, and other SM particles, one could in principle search for them as point-like sources.  However, due to the small horizon radius for a light PeBH, the electromagnetic radiation power has $\sim 2\times 10^{-10}\,\mbox{W}\times (M_{\rm eBH}/10^9\mbox{g})^4$, which is too weak to be observed as a single source. Other than single source searches, the Hawking radiation of PeBH's can also contribute to the diffuse photon flux. For a Boltzmann energy spectrum with a temperature $T_{\rm eBH}$, the averaged energy of the radiated photons has $\langle E_\gamma \rangle \approx 2.7\, T_{\rm eBH}$. The photon spectral emission rate from one PeBH has 
\beqa
\frac{dN_\gamma}{d E_\gamma \,d t} \approx \frac{\pi^2}{60}\, 4\pi\,r_+^2 \, T_{\rm eBH}^4 \,\frac{1}{\langle E_\gamma \rangle^2} \approx \frac{0.54\,M_{\rm eBH}^3}{\Mpl^4\,t_{\rm univ}} ~,
\label{eq:emssion-rate}
\eeqa
which has a higher rate for a heavier black hole mass. 

Within our galaxy and similar to a decaying DM model, one can adopt the normalized ``$J$"-factor after integrating along the line of sight to calculate the PeBH radiated flux
\beqa
\widehat{J} \equiv \frac{1}{4\pi\,M^{\rm total}_{\rm DM} }\int \cos{b}\,\mbox{d}b\,\mbox{d}\ell \,\mbox{d} s \, \rho_{\rm DM}\left[ r(s, \ell, b)\right] \,,
\eeqa
with $M^{\rm total}_{\rm DM} \equiv 4\pi\,\int \mbox{d}r\,r^2\,\rho_{\rm DM}(r)$. Here, $-\pi/2 \leq b < \pi/2$ and $0 \leq \ell < 2\pi$ are galactic latitude and longitudinal angles; $s$ is the line-of-sight distance with $r^2 = s^2 \cos^2{b} + R_\odot^2 - 2 s\,R_\odot \cos{\ell} \cos{b}$ and the distance of the Sun to the galactic center as $R_\odot \approx 8.5$\,kpc.   Using the Einasto profile with $\beta=0.17$~\cite{Graham:2006ae}, the solid-angle-averaged  $J$-factor has $\widehat{J} \approx 1/(33\,\mbox{kpc})^2\,\mbox{sr}^{-1}$. Combining with \eqref{eq:emssion-rate}, the contribution to the sky-averaged diffuse photon spectral flux is estimated by
\beqa
\phi_\gamma&\approx& \frac{M^{\rm total}_{\rm DM} }{M_{\rm eBH}}  \widehat{J} \frac{dN_\gamma}{d E_\gamma \,d t} \,  \nonumber \\
&\approx& 9\,\times 10^{-11}\,\mbox{cm}^{-2}\mbox{sr}^{-1}\mbox{s}^{-1}\mbox{MeV}^{-1}  \times \left( \frac{M_{\rm eBH}}{10^9\mbox{g}}\right)^2\,, \nonumber 
\eeqa
for the averaged photon energy of $\langle E_\gamma \rangle \approx 0.4\,\mbox{MeV} \,\times\, [M_{\rm eBH}/(10^9\,\mbox{g})]^{1/2}$. Comparing to the spectra of isotropic diffuse gamma ray measured by the COMPTEL collaboration~\cite{COMPTEL} and the EGRET data~\cite{Sreekumar:1997un,Strong:2004ry}, the PeBH mass is constrained to be $M_{\rm eBH}\lesssim 10^{12}$\,g assuming that they compose nearly all or all of DM. 

Big Bang Nucleosynthesis (BBN) and the Cosmic Microwave Background (CMB) can also constrain both stages of PeBH evaporation during the early universe. For the generic case with $\Minit \gg M_\text{eBH}$, the evaporation of the PBH towards PeBH may generate enough radiation energy to affect BBN or CMB observables. This is true for $\Minit \gtrsim 1.4 \times 10^{9}$~g, or equivalently $\tau_{\rm BH} \gtrsim 1~\text{s}$ when BBN starts. Because the PeBH is lighter than its initial BH mass, the constraints on the fraction of PeBHs making up DM, $f_{\rm eBH}$, should be more stringent by a factor of $\Minit/M_\text{eBH}$ compared to the existing constraints~\cite{Carr:2009jm} on ordinary PBHs. On the other hand, for $\Minit \lesssim 1.4 \times 10^{9}$~g, the initial BH has already reached the near-extremal state before BBN and there is no bound from the initial evaporation. 

Emissions by BHs that have already reached near-extremal are also constrained, albeit more weakly owing to their reduced temperature.
For example, if the Hawking temperature is high enough to produce hadrons, it will affect the light-element abundances. 
Rather than redo the detailed analysis, we make use of previous results to estimate the BBN bounds in the following way. For ordinary PBHs, the effects of hadron injection depend on the hadron emission rate $\mbox{B}_{\rm h}\,n_{\rm BH}\,\langle{E}(t)\rangle^{-1} \, dM_{\rm BH}/dt$ with $\mbox{B}_{\rm h} = \mathcal{O}(1)$ as the hadronic branching ratio and $\langle{E}(t)\rangle \sim T_{\rm BH}(t)$~\cite{Kohri:1999ex}. The most stringent constraint comes from the Lithium-6/Lithium-7 abundance ratio with $f_{\rm BH} \lesssim 6.8\times 10^{-4}$ for $M_{\rm BH} \approx 4\times 10^{10}$~g or $\tau_{\rm BH} \approx 2.6\times 10^4$~s, which we use as a reference point to derive the constraints on PeBHs. The constraint on $f_{\rm eBH}$ can be derived to be $f_{\rm eBH} \lesssim 4.2 \times 10^{-3}\, \times\,[(4 \times 10^{10}\, \text{g})/M_{\rm eBH}]^{5/2}$. In practice, this constraint is generally superseded by constraints on the earlier evaporation from $\Minit$.

\begin{figure}[tb!]
\centering
\includegraphics[width=0.48\textwidth]{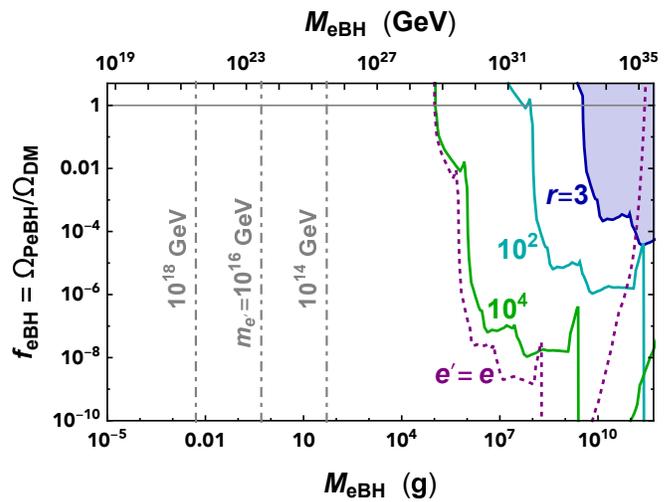}
\caption{The BBN and other black hole evaporation-related constraints on the PeBH energy density as a function of PeBH masses for various ratios of the initial BH mass over the PeBH mass $r \equiv \Minit/M_\text{eBH}$. The purple dotted line shows the constraints for one possible PeBH cosmological formation scenario described around \eqref{eq:Ncol} with $e'=e$. In the vertical and gray dot-dashed lines, the potential theoretical constraints from Schwinger evaporation and the WGC via \eqref{eq:Mmin-WGC} are shown for different dark electron masses.  
}
\label{fig:BBN-constraint}
\end{figure}

Scaling the constraints for ordinary PBHs~\cite{Carr:2009jm}, we show the constraints on the fraction of PeBHs as DM in Fig.~\ref{fig:BBN-constraint} for several values of $r \equiv \Minit/M_\text{eBH}$. The jumps in the exclusion curves relate to different element abundance ratios; the largest jump is where the CMB constraints become important. The constraint from the Hawking emission of the near-extremal BHs (as opposed to the Hawking radiation from the initial BHs) is only important in Fig.~\ref{fig:BBN-constraint} in the left-most portion of the $r=3$ curve. One can see from Fig.~\ref{fig:BBN-constraint} that there is a wide range of open parameter space (from the Planck scale to $\sim 10^9$\,g) to have PeBHs account for 100\% of DM. We also show in the vertical and gray dot-dashed lines the combined theoretical constraints from Schwinger evaporation together with the WGC from Eq.~\eqref{eq:Mmin-WGC}. For a lighter dark electron mass, the lower bound on the PeBH mass becomes larger.  We note that if the WGC is not valid, one can relax the Schwinger-evaporation constraints by choosing a tiny gauge coupling $e'$ [see Eq.~\eqref{eq:Mmin}]. If both $\mep$ and $M_{\rm eBH}$ can be measured, one could test the WGC experimentally. While not displayed on the plot, when $r \rightarrow 1$ (meaning the black hole was born very near extremal), the BBN constraints are dramatically relaxed by a factor of $[(\Minit -M_\text{eBH})/M_\text{eBH}]^{3/2}$.~\footnote{We regard the possibility that the PBHs are born in an already near-extremal state as less likely based on the formation mechanism presented later.  However, if such a formation model is realized, its evaporation would be less constraining.}

%

\section{Signatures of binary $\mbox{PeBH}$ mergers}
Other than signatures from isolated PeBHs, one could also search for signatures from binary PeBH systems. Although the merger event of a PeBH binary could generate gravitational wave signatures, the signal flux is too small to be detected for a lighter PeBH mass far below the solar mass. On the other hand, the PeBH binary system typically contains both positively and negatively charged black holes (like-charged PeBHs have no net force to bind them). When they merge, the charges are neutralized such that the final black hole becomes non-extremal and can provide striking Hawking radiation signatures.

Given the large uncertainties on estimating the black hole binary merging rate~\cite{Raidal:2018bbj}, which are most severe when $f_\text{eBH} = \mathcal{O}(1)$ and come predominantly from binary disruption both immediately after binary formation and during halo formation, we simply make a phenomenological parametrization for the rate and discuss some benchmark-model estimation at the end of this section. For an approximately monochromatic PeBH mass distribution with mass $M_{\rm eBH}$, the averaged binary merging rate with equal BH masses can be parametrized as
\beqa
\Gamma_{\rm merge} = \Gamma_0 \, f_{\rm eBH}^a\, \left( \frac{M_{\rm eBH}}{M_{\rm ref}} \right)^b \, \left(\frac{t}{t_{\rm univ}} \right)^c \,,
\eeqa
where the exponents $a, b, c$ depend on the binary formation history. Ignoring the disruption effects in the galaxy formation, the merging rate per volume in the current Milky-Way galaxy is
\beqa
R(r) = \frac{\rho_{\rm DM}(r)}{4\, M_{\rm eBH}} \, \Gamma_0 \, f_{\rm eBH}^{a+1}\,\left( \frac{M_{\rm eBH}}{M_{\rm ref}} \right)^b \,, 
\eeqa
with one factor of 1/2 to avoid overcounting, and the second factor of 1/2 because only oppositely charged PeBHs form binaries.

After two $+Q$ and $-Q$ PeBH's merge and ignoring the angular momentum, they become approximately an ordinary black hole with a mass of $M_{\rm BH} \approx 2 M_{\rm eBH}$. As the BH evaporates,  the Hawking temperature, $T_{\rm BH} \approx 53\,\mbox{TeV}\times(10^8\,\mbox{g}/M_{\rm eBH})$, is much higher than that of an isolated PeBH. The duration of the evaporation is very short: $\tau_{\rm BH} \approx 10^{-3}\,\mbox{s}\,\times (M_{\rm eBH}/10^8\,\mbox{g})^3$. So, the final black hole evaporation is a very transient event. 

Both high-energy neutrinos and gamma rays are good signatures to search for the final evaporation of the merged binary system. Starting with the high-energy neutrinos, the number of neutrinos from one evaporation event is estimated to be $\eta_\nu\,(2 M_{\rm eBH})/\langle E_\nu\rangle$ with the averaged neutrino energy $\langle E_\nu\rangle \approx 4.22\, T_{\rm BH}\approx 220\,\mbox{TeV}\times(10^8\,\mbox{g}/M_{\rm eBH})$~\cite{Carr:2009jm} and $\eta_\nu \approx 0.06$ as the fraction of energy into primary three-flavor neutrinos after taking into account the different thermally averaged cross sections~\cite{Page:1976df}. Including the averaged binary merger rate, the neutrino emission rate is estimated to be 
\beqa
\frac{dN_\nu}{dt} \approx1.4\, \frac{M_{\rm eBH}^2}{\Mpl^2} \,\Gamma_{\rm merge} ~. 
\eeqa
This only accounts for the primary neutrinos, ignoring secondary neutrinos produced in the cascade decays of other Hawking-emitted particles. Using the normalized $J$-factor of the Einasto profile, the predicted high-energy neutrino flux originating from the Milky Way halo is
\beqa
\Phi_\nu &\approx& \frac{M^{\rm total}_{\rm DM} }{4 \, M_{\rm eBH}} \, \widehat{J}\, \frac{dN_\nu}{dt}   \\
&& \hspace{-0.9cm} \approx~ 5 \times 10^{-15}\mbox{cm}^{-2}\mbox{sr}^{-1}\mbox{s}^{-1} \left( \frac{M_{\rm eBH}}{10^8\,\mbox{g}}\right)^{1+b} \left(\frac{f^{a+1}_{\rm eBH}\,\Gamma_0}{10^{-30}\,\mbox{s}^{-1}}\right)  \,. \nonumber 
\eeqa
The extragalactic flux may be comparable to or a couple of orders of magnitude smaller than the galactic flux---see, {\it e.g.},~\cite{Ahlers:2015moa}. 
Note this treatment makes the important but unqualified assumption that the binary merger distribution identically follows the DM distribution.  This is less likely to be true within DM halos, although detailed study is needed.

Proceeding despite this, the lack of an observation of neutrinos above the PeV scale by IceCube~\cite{Aartsen:2018vtx} along with constraints from ANITA~\cite{Gorham:2019guw} and NuMoon~\cite{Scholten:2009ad,Buitink:2010qn} limit $\Phi_\nu \lesssim (2.7\times 10^{-15}~\text{cm}^{-2}\text{s}^{-1}\text{sr}^{-1})(E_\nu/10^7~\text{GeV})^{-4/5}$ for $10^7~\text{GeV} \lesssim E_\nu \lesssim 10^{16}~\text{GeV}$. 
Further, the observation of cosmic neutrinos by IceCube with energies $10^4~\text{GeV} \lesssim E_\nu \lesssim 10^7~\text{GeV}$~\cite{Aartsen:2018fqi} can be used to set an upper limit $\Phi_\nu \lesssim (5.9 \times 10^{-16}~\text{cm}^{-2}\text{s}^{-1}\text{sr}^{-1})(E_\nu/10^7~\text{GeV})^{-8/5}$.
This leads to a constraint on the merging rate of roughly~\footnote{As a cross-check, this bound is fairly similar to the lifetime constraint on decaying DM made by IceCube~\cite{Aartsen:2018mxl}, though the latter included both halo and extragalactic components as well as a fuller accounting of all neutrino production channels.}
\beqa
\label{eq:merge-rate-constraint}
&&  \hspace{0mm}\Gamma_0\,f^{a+1}_{\rm eBH}\lesssim  \\
&&  \hspace{-0mm} \Bigg\{
\begin{array}{l r}
 \left(10^{-29}\,\mbox{s}^{-1}\right) \left( \dfrac{10^8\,\mbox{g}}{M_{\rm eBH}}\right)^{{1\over5}+b} \,, \hspace{6mm}  M_\text{eBH} \in (10^{-3}, 10^6)\,\mbox{g}  \,,
 \\
\left(5 \times 10^{-29}\,\mbox{s}^{-1}\right) \left( \dfrac{10^8\,\mbox{g}}{M_{\rm eBH}}\right)^{b-{3\over5}} \,, \hspace{1mm}   M_\text{eBH} \in (10^{6}, 10^9)\,\mbox{g} \,. \nonumber 
\end{array}
\eeqa
Comparing against Fig.~\ref{fig:BBN-constraint}, this bound extends down to the lower limits on $M_\text{eBH}$ coming from the WGC, though in principle one could search for still higher-energy neutrinos to get down to $M_\text{eBH} \sim \Mpl$.
On the other side, this bound extends all the way up to PeBH masses that are generally constrained by the evaporation from $\Minit$, though in principle bounds on binary mergers could be extended to higher $M_\text{eBH}$ using gamma ray bounds from HAWC \cite{Pretz:2015wma,Abeysekara:2017jxs} and CTA \cite{Acharya:2017ttl}.

Other than searching for only high-energy neutrinos or only gamma rays, one could also use the multi-messenger approach to search for coincident cosmic ray events by several telescopes: IceCube, ANITA, NuMoon, HAWC, CTA, Fermi-LAT \cite{Ackermann:2012rg,Ackermann:2014usa}, Pierre Auger~\cite{Zas:2017xdj}, and others~\cite{Tesic:2015gpk,galaxies7010019}.  This would be a particularly powerful discovery channel that can discriminate against other high-energy backgrounds and allow us to directly detect Hawking radiation.

An estimate for the merger rate is derived in the appendix which has $a=1/3$, $b=1/9$, $c=-8/9$, and $\Gamma_0 \approx 1.5\times 10^{-23}~\text{s}^{-1}$ for $M_\text{ref}=10^{8}~\text{g}$ (with a monochromatic mass distribution) and numerical factors $\alpha=\beta=1$ [see Eqs.~(\ref{eq:xbar}-\ref{eq:GammaMergeEM})].  This estimate assumes that the PeBHs in the to-be-formed binary decouple from Hubble expansion shortly before matter-radiation equality and form long-lived binaries due to the tidal force of neighboring PeBHs. Na\"ively, it seems to be very constrained by (\ref{eq:merge-rate-constraint}). However, as argued in Ref.~\cite{Raidal:2018bbj}, these binaries may be disrupted in the early universe when $f_{\rm eBH} = \mathcal{O}(1)$, resulting in a lengthened merger time and a dramatically decreased merger rate---by their estimate, a decrease of up to but not larger than a factor of $10^{17}$ for $M=10^8~\text{g}$ PBHs.~\footnote{Their analysis only considered gravitation.  Besides accounting for the change in merger time, properly adding dQED would require a completely separate N-body simulation because same-charged PeBHs do not interact.}
Thus, we expect that our estimation for $\Gamma_0$ may severely overestimate the merger rate.

Additional disruptions during halo formation may further suppress the merger rate in the Milky Way halo.  Since the extragalactic binary population is more stable against these disruptions, their mergers may provide more robust constraints on PeBHs if the extragalactic flux is similar to or within a couple of orders of magnitude of the undisrupted galactic flux~\cite{Ahlers:2015moa}.  Further work is needed to understand the halo disruptions.  On the other hand, a merger occurring nearby in the Milky Way halo or one of its dwarf galaxies may be more likely to be detected with several instruments observing different messengers.  Thus, nearby mergers may offer better discovery prospects.

\section{$\mbox{PeBH}$ formation}
So far, we have remained agnostic to how the population of PeBHs is formed and focused instead on what the phenomenological consequences of such a population would be.  Here, let us briefly highlight one possibility for their formation relying on the statistical distribution of charges in the early universe.

Assume that the PBHs are initially formed at some earlier time corresponding to an initial black hole mass $\Minit$ during radiation domination following reheating. 
The details for this formation are not important and could be the result of any number of already or as-yet proposed inflationary or post-inflationary models~\cite{Carr:1974nx,Carr:2016drx,Orlofsky:2016vbd,Sasaki:2018dmp}.  We simply assume that they form as a result of primordial matter overdensities that collapse almost immediately after reentering the horizon. We will further assume that the dark electrons are still relativistic at some temperature $T'$ when the PBHs form. Note that the dark sector temperature $T'$ may not be equal to the visible sector temperature $T$.
Since the dark photon is massless, the constraints on additional radiation degrees of freedom from CMB measurement require $T'/T \lesssim 0.46$~\cite{Aghanim:2018eyx}.

One may then ask what the dark charge of these to-be-collapsed regions are as they reenter.  We take the ansatz that reheating occurs in such a way that the dark electrons and positrons each follow independent Poisson distributions in superhorizon volumes. Thus, the average number of each within a volume that is just reentering the horizon is $\left<N_\text{col}\right> \approx \frac{4\pi}{3} n_{e'} H^{-3}$, with $H=[(8\pi/3) \rho_r/\Mpl^2]^{1/2}$ the Hubble parameter and $\rho_r$ the radiation energy density.  Noting that $\Minit \approx \frac{4\pi}{3}\,\rho_r\,H^{-3}$, the average number of a particular species (either dark electron or positron) within the collapsed volume is
\beqa
\label{eq:Ncol}
\langle N_{\rm col} \rangle
&\approx& 3 \times 10^{20} \times  \left(\frac{\Minit}{10^9\,\mbox{g}} \right)^{3/2} \left(\frac{T'}{T} \right)^3  \,.
\eeqa
If the dark electron and positron distributions are indeed Poisson and independent, the resulting charge distribution of the PBHs is a Skellam distribution~\cite{10.2307/2981372} with mean $\left<Q\right>=0$ and standard deviation $\sigma_{Q,\text{col}} = \sqrt{2 \left<N_\text{col}\right>}$. 

After the formation of the initial non-extremal black hole with a charge $Q$, the ambient dark electrons and positrons can be further absorbed by black holes and change the charge distribution. In the early stage before the PBHs have evaporated to near-extremal, the gravitational interaction can be assumed under suitable model parameters to be stronger than the dark electromagnetic interaction, so one can approximately ignore the rate difference for dark electron and positron absorption. The net effects would be an increase of the variance $\sigma_Q$ and a slight change of the charge distribution. More quantitatively, starting from the time that the BH is formed with mass $\Minit$, the non-relativistic PBH will encounter and absorb relativistic dark electrons or positrons at a rate $n_{e'} \sigma v$. Compared to the Hubble scale, the absorption rate is efficient as long as
\beqa
\frac{n_{e'} \sigma v}{H} \approx 0.82\,\frac{g}{\sqrt{g_*}} \left(\frac{T'}{T}\right)^3 \left(\frac{T\,M^2}{\Mpl^3}\right) \gtrsim 1 ~,
\eeqa
where $g$ is the number of degrees of freedom for $e'$ and we have approximated $\sigma \approx \pi r_+^2 \approx \pi (2\,G_N M)^2$ when $M \gg M_\text{eBH}$ (when $M \approx M_\text{eBH}$, $r_+ \approx G_NM$). Here, we use the geometric cross section, which is valid when the wavelength of $e'$ is shorter than the black hole radius~\cite{Hawking:1971ei}, {\it i.e.}, $T' \gtrsim \Mpl^2/(2 M) \approx 10^5~\text{GeV} \times (10^9~\text{g}/M)$ assuming $T'>\mep$.

This absorption process may freeze out in a few different ways: $i)$ the number density of dark electrons and positrons decreases ({\it e.g.}, after $T' < \mep$), $ii)$ the PBH evaporates to a PeBH and its geometric cross section shrinks, or $iii)$ the dark electromagnetic force becomes important in repelling like-charged particles and attracting unlike-charged particles.  Note that $(iii)$ occurs automatically once the PBH evaporates to a PeBH if both $n_{e'}$ and the geometric cross section are not too small and if the WGC holds.  Since $(iii)$ can lead to discharge, we will generally prefer that freeze-out occurs in a different way. This can be arranged, {\it e.g.}, by having $T'$ drop below $\mep$ and allowing the dark electrons/positrons to annihilate away before the PBH has time to evaporate, satisfying $(i)$. Then, after the black hole evaporates to (nearly) extremal, the charge neutralization rate can remain smaller than the Hubble rate, so the PeBH with a charge $Q$ can survive until the current universe.  Let us estimate the total charge under those assumptions.

It turns out that if the PBH formation temperature is much greater than the freeze-out temperature, then most of the absorption occurs near the formation temperature.  Still assuming that the PBH formed during radiation domination, the number of dark electrons and positrons absorbed can then be estimated by integrating $n_{e'} \sigma v$ as
\beqa
\hspace{-3mm} N_\text{absorp} \approx 4 \times 10^{18} \left(\frac{\Minit}{10^{9}~\text{g}} \right)^{3/2} \left(\frac{T'}{T} \right)^3 \left(\frac{g_*}{100} \right)^{-3/4}.
\eeqa
This results in a one-dimensional random walk for the charge accumulation with $N_\text{absorp}$ steps, which is characterized by a standard deviation of $\sigma_{Q,\text{absorp}} = \sqrt{N_\text{absorp}}$.  While it is subdominant to the charge from the initial collapse in \eqref{eq:Ncol}, it adds to the charge distribution to give a total expected charge of about $\sigma_Q = \sqrt{2 N_\text{col} + N_\text{absorp}}$.

In the case without too much discharge, then we may expect a typical PeBH mass of $M_\text{eBH} \approx e' \sigma_Q \Mpl$ if it is above the Schwinger-evaporation bound in Eq.~\eqref{eq:Mmin}; otherwise if it is below the evaporation bound, the surviving PeBH masses will be peaked just slightly above $M^{\rm min}_{\rm eBH}$.  Constraints assuming negligible discharge are shown in Fig.~\ref{fig:BBN-constraint} for $e'=e$, where PBHs form with $r \approx 10^4$ at smaller $M_\text{eBH}$ and form with even larger mass ratio with increasing $M_\text{eBH}$.~\footnote{At the far right of the plot in Fig.~\ref{fig:BBN-constraint}, $r$ is so large that the initial BH is cosmologically stable and does not evaporate to near-extremal.}

\section{Discussion and conclusions}
In the model considered here to stabilize light PeBHs against discharge, the dark electron is assumed to be the lightest dQED-charged particle and hence stable. In our analysis, we have assumed that its contribution to the DM energy density is subdominant compared to the PeBH contribution. However, if the very heavy dark electron has a thermal relic abundance set by its annihilation into dark photons, it will overclose the universe. This issue can be resolved by introducing additional non-Abelian low-scale confining gauge charges for the dark electrons (see, {\it e.g.}, Refs.~\cite{Kang:2006yd,Jacoby:2007nw}). 

So far, we have only considered gravitational interaction between the visible and dark sector. Additional interactions like kinetic mixing between the photon and dark photon~\cite{Holdom:1985ag} can make dark electrons and PeBHs millicharged under ordinary electromagnetism. However, in order that Schwinger pair production of SM electrons be inefficient, the mixing parameter is required from Eq.~(\ref{eq:Mmin}) to be $\epsilon \lesssim 1.6 \times 10^{-19}$ (independent of the small $M_\text{eBH}$ considered here).  Because the PeBH's SM electromagnetic charge is $(\epsilon\, e/e') \sqrt{4\pi} M_\text{eBH}/\Mpl$, such a small mixing would be hard to detect experimentally unless $e'$ was much smaller than allowed by the WGC.  Note that while such a small $\epsilon$ is allowed from a low-energy perspective, its value may be restricted by the details of the high-scale theory (see, {\it e.g.}, \cite{Dienes:1996zr,Abel:2008ai,Goodsell:2009xc}).
Another possibility reserved for future work is to replace $U(1)_\text{dark}$ by the $U(1)_{\rm B-L}$ gauge group with a tiny gauge coupling and neutrinos as the lightest charged particles~\cite{Cheung:2014vva}. 

In summary, we have studied the possibility of having PeBHs as a DM candidate based on a dQED model with a heavy dark electron. A wide range of PeBH masses from the Planck scale to around $10^9$~g are still allowed by the experimental constraints. The merger event of two equal-mass PeBHs with opposite charges can generate a neutral and light Schwarzschild black hole whose fast Hawking evaporation can provide transient signatures to be detected by various telescopes. While a simple early-universe formation story for PeBHs has been discussed in this paper, other formation mechanisms based on different histories are worthy of future studies.

\vspace{2mm}
The work is supported by the U. S. Department of Energy under the contract DE-SC0017647. Part of this work was performed at the Aspen Center for Physics, which is supported by National Science Foundation grant PHY-1066293.

\vspace{2mm}


\appendix
\section{Appendix I: binary merger rate}
\label{appendix-I}
The PeBHs may attract each other via gravitational and dQED forces to form binaries in the early universe.  Let us take the simplifying assumption that PeBHs form during radiation domination, and the first period of matter domination is the ordinary one occurring at redshift $z_\text{eq} \approx 3400$ \cite{Aghanim:2018eyx}.  In order to decouple from Hubble expansion, a pair of PeBHs must dominate their local energy density. Thus, such binaries only form when $z>z_\text{eq}$.~\footnote{The merger rate of binaries formed in galactic halos is subdominant.  For example, using the estimate in Ref.~\cite{Ali-Haimoud:2017rtz} for the purely gravitational case, the halo-formed binary merger rate is about five orders of magnitude smaller for $M_\text{eBH}=10^8~\text{g}$.}  They may subsequently merge, and their merger rate can be estimated.

To calculate the merger rate, we update the treatment in \cite{Ioka:1998nz,Sasaki:2016jop} (although see further refinements in \cite{Ali-Haimoud:2017rtz,Raidal:2018bbj}) to account for radiation due to dQED, which dominates the gravitational radiation.  First, the average separation at matter-radiation equality is
\begin{equation}
\label{eq:xbar}
\bar{x} = \left(\frac{M_\text{eBH}}{\rho_\text{PeBH}(z_\text{eq})}\right)^{1/3} = \frac{f_\text{eBH}^{-1/3}}{(1+z_\text{eq})} \left(\frac{8 \pi M_\text{eBH}}{3 \Mpl^2 H_0^2\,\Omega_\text{DM}}\right)^{1/3}.
\end{equation}
A pair of nearest-neighbor PeBHs form into a binary if $x<f_\text{eBH}^{1/3}\bar{x}$ depending on the distance to the next-nearest-neighbor PeBH $y$ (which provides a tidal force to prevent the other two PeBHs from immediately merging).  The resulting orbital parameters are~\cite{Ioka:1998nz}
\begin{equation}
a = \frac{\alpha}{f_\text{eBH}} \frac{x^4}{\bar{x}^3}~,
\; \; \; \; \;
b = \beta \left(\frac{x}{y}\right)^3 a~,
\; \; \; \; \;
e = \sqrt{1 - \left(\frac{b}{a}\right)^2} ~,
\end{equation}
indicating the semi-major axis, semi-minor axis, and eccentricity, respectively, with $\alpha$ and $\beta=\mathcal{O}(1)$ numbers that should be determined through numerical simulations.  Then, the merger time for two PeBHs with masses $M_\text{eBH}=M_{1,2}$ is [see the detailed derivation in the next section leading to \eqref{eq:merging-time-formula}]
\beqa
\label{eq:mergertime}
t &=& \frac{1}{16} \frac{\Mpl^4}{M_1 M_2} a^3 \frac{(1-e^2)^{5/2}}{2+e^2} 
 \equiv 
\bar{t} f_\text{eBH}^{3} \left(\frac{a}{\bar{x}}\right)^3 \frac{(1-e^2)^{5/2}}{(2+e^2)/3} 
\nonumber
\\
& \approx& \alpha^3 \beta^5 \bar{t} \left(\frac{x}{\bar{x}}\right)^{27} \left(\frac{y}{\bar{x}}\right)^{-15}  ~,
\eeqa
where in the last line we approximate $e \approx 1$, which is valid for mergers occurring today.  Here, the time scale $\bar{t} \equiv \Mpl^4 \,\bar{x}^3/(48 f_{\rm eBH}^3 M_1 M_2)$. This can be compared to the merger time for ordinary, uncharged PBHs emitting only gravitational radiation for $e \approx 1$~\cite{Peters:1964zz}
\beqa
t_\text{grav} = \frac{3}{85}\,\frac{\Mpl^6}{M_1M_2(M_1+M_2)}\,a^4\,(1-e^2)^{7/2} ~.
\eeqa
Taking $M_1=M_2=M$, their ratio is
\beqa
\label{eq:mergertimeratio}
\frac{t_\text{grav}}{t} & = & \frac{72}{85} \frac{\Mpl^{2}}{M} a (1-e^2)\frac{2+e^2}{3}
\\
& \approx& 2 \times 10^{14} \left(\frac{10^9~\text{g}}{M}\right) \left(\frac{a}{f^{1/3} \bar{x}}\right) \left(\frac{1-e^2}{10^{-14}}\right) \frac{2+e^2}{3}  ~, \nonumber 
\eeqa
where fiducial values of $a$ (with $\bar{x}$ evaluated using $M=10^9~\text{g}$) and $1-e^2$ are chosen so that $t \approx t_\text{univ}$ (the full probability distribution for these is considered below).  As expected, the dQED radiation makes the binary merge more quickly than gravity alone.

The normalized probability distribution is given by~\cite{Ioka:1998nz}
\beqa
dP = \frac{9 x^2 y^2}{\bar{x}^6} e^{-(y/\bar{x})^3} dx dy  ~.
\eeqa
After changing variables from $x$ to $t$ using \eqref{eq:mergertime} and properly taking care of the integration region of $y$, we integrate $0<x<f_\text{eBH}^{1/3}\bar{x}, \, x<y<\infty$ to arrive at the merging rate as
\beqa
\label{eq:GammaMergeEM}
\Gamma_\text{merge} &= & \frac{dP}{dt} = \frac{1}{9} \left(\frac{t}{ \alpha^3 \beta^5\bar{t}}\right)^{1/9} \frac{1}{t}  \times  \\
&& \hspace{-1.4cm} \left\{ \Gamma\left[\frac{14}{9}, \left(\frac{t}{ \alpha^3 \beta^5 \bar{t}}\right)^{1/4}\right]  -   \Gamma\left[\frac{14}{9}, \left(\frac{t}{ \alpha^3 \beta^5 \bar{t}}\right)^{-1/5} f_\text{eBH}^{9/5}\right]  
\right\},
\nonumber
\eeqa
with $\Gamma[i, z]$ as the incomplete gamma function. 

The merger rate per unit volume is then,
\beqa
R = \frac{n_\text{PeBH}}{4} \frac{dP}{dt} (t=t_{\rm univ}) ~,
\eeqa
where one factor of $1/2$ is to avoid overcounting, and the second factor is to account for the fact that only oppositely-charged PeBHs form binaries.
It is often sufficient to approximate the $\Gamma$ function as a constant, in which case
\beqa
R \propto f_\text{eBH}^{13/9} M_\text{eBH}^{-8/9} t^{-8/9} ~.
\eeqa
Note, this equation matches the general structure of \eqref{eq:merge-rate-constraint}. By constrast, for uncharged PBHs, $R \propto f_\text{eBH}^{53/37} M^{-32/37} t^{-32/37}$.  While the exponents are numerically similar, the overall normalization has the merger rate for charged PeBHs about two orders of magnitude larger than that of ordinary PBHs for the lighter masses of interest here [note that the lifetime only enters as $\bar{t}^{-1/9}$ in $\Gamma_\text{merge}$ in (\ref{eq:GammaMergeEM}), and a similarly small exponent for the gravitational case, which explains the relationship to (\ref{eq:mergertimeratio})].

\section{Appendix II: dQED radiation}
To calculate the merger time for PeBH binaries, consider a central inverse-square force acting on a body with mass $M_1$ and charge $Q_1$ due to another body with mass $M_2$ and opposite charge $Q_2$ satisfying
\beqa
F_{12} = -2\, \frac{e'^2}{4 \pi} \frac{|Q_1 Q_2|}{r^2} ~,
\eeqa
where $r = |\bm{x}_1-\bm{x}_2|$ is the distance between the two PeBHs.
The factor of 2 comes from the fact that we have assumed these BHs are extremal and oppositely charged, so the force of gravity and of dQED are equal and add constructively.

These charges form a time-varying dark electric dipole
${\bm{p}} = e' Q_1 \bm{x}_1 + e' Q_2 \bm{x}_2$
that emits radiation as
\begin{equation}
\frac{dE}{dt} = -\frac{2}{3} \frac{1}{4 \pi} \,\ddot{p}^2 = -\frac{2}{3} \frac{\gamma^2}{4 \pi}  \frac{1}{r^4},
\end{equation}
where $\gamma \equiv 2 e'^3 \left(|Q_1|/M_1 + |Q_2|/M_2\right) |Q_1 Q_2| / (4 \pi)$.

Using the standard expression for the separation between two orbiting bodies as a function of the semimajor axis $a$, eccentricity $e$, and angle that $\bm{r}$ makes with the periapsis $\theta$,
\begin{equation}
r = \frac{a (1-e^2)}{1+e \cos \theta} ~,
\end{equation}
as well as the eccentric anomaly to parametrize the motion as a function of time:
\beqa
&& t = \frac{T}{2\pi} (\psi - e \sin \psi)  ~, \nonumber 
\\
&& (1-e \cos\psi)(1+e \cos\theta) = 1-e^2 ~, \nonumber 
\eeqa
the average energy loss over an orbital period $T$ is,
\beqa
\left<\frac{dE}{dt}\right> & \approx& -\frac{2}{3} \frac{\gamma^2}{4 \pi } \frac{1}{T} \int_{0}^{T} \frac{1}{r^4} \, dt  \nonumber 
\\
& =& -\frac{2}{3} \frac{\gamma^2}{4 \pi} \frac{1}{a^4} \frac{1}{2 \pi} \int_{0}^{2 \pi} (1-e\cos\psi)^{-3} d\psi   \nonumber 
\\
& =& - \frac{1}{3} \frac{\gamma^2}{4 \pi } \frac{1}{a^4} \frac{2+e^2}{(1-e^2)^{5/2}} ~.
\eeqa
This can be related to the total energy
\beqa
E = -2 \frac{e'^2 \,|Q_1 Q_2|}{(4 \pi) (2a)} \equiv - \frac{\gamma'}{2a}  ~, 
\eeqa
by the expression
\beqa
\left<\frac{dE}{dt}\right> & \approx& - \frac{16}{3} \frac{1}{4 \pi } \frac{\gamma^2}{\gamma'^4} E^4 \frac{2+e^2}{(1-e^2)^{5/2}}  ~. \nonumber
\eeqa
Then, one can integrate $E$ from some initial $E(a_i)$ to $-\infty$ to calculate the merger time as
\beqa
t & \approx& \frac{1}{2} \frac{4 \pi \gamma'}{\gamma^2}a^3 \frac{(1-e^2)^{5/2}}{2+e^2} \nonumber
\\
& =& \frac{4\pi^2 a^3}{e'^4 |Q_1 Q_2| (|Q_1|/M_1 + |Q_2|/M_2)^2}  \frac{(1-e^2)^{5/2}}{2+e^2}   ~,
\label{eq:merging-time-formula}
\eeqa
where we have dropped the subscript $i$ in $a_i$.  Finally, one can substitute $e'|Q|= \sqrt{4 \pi}M/\Mpl$ to obtain (\ref{eq:mergertime}).

\bibliographystyle{apsrev4-1}
\bibliography{extremal_pbh_refs}
\end{document}